\documentclass{iaus}
\usepackage{graphicx}
\usepackage{natbib}

\title[Galactic star formation at parsec-scales] 
{Galactic star formation in parsec-scale resolution simulations}

\author[Leila C. Powell et al]   
{Leila C. Powell$^1$, Frederic Bournaud$^1$, Damien Chapon$^1$, Julien Devriendt$^2$, Adrianne Slyz$^2$ and Romain Teyssier $^{1,3}$}

\affiliation{$^1$ SAp, CEA-Saclay, F-91191 Gif-sur-Yvette Cedex, France \\email: {\tt leila.powell@cea.fr}\\ $^2$ Oxford Astrophysics, Denys Wilkinson Building, Keble Road, Oxford, OX1 3RH, UK\\ $^3$ Institute of Theoretical Physics, University of Zurich Winterhurerstrasse 190, CH-8057 Zurich, Switzerland}
\pubyear{2010}
\volume{270}  
\pagerange{xxx--xxx}
\jname{Computational star formation}
\editors{Alves, Elmegreen, Girart, Trimble, eds.}
\begin{document}

\maketitle

\begin{abstract}

The interstellar medium (ISM) in galaxies is multiphase and cloudy, with stars forming in the very dense, cold gas found in Giant Molecular Clouds (GMCs). Simulating the evolution of an entire galaxy, however, is a computational problem which covers many orders of magnitude, so many simulations cannot reach densities high enough or temperatures low enough to resolve this multiphase nature. Therefore, the formation of GMCs is not captured and the resulting gas distribution is smooth, contrary to observations. We investigate how star formation (SF) proceeds in simulated galaxies when we obtain parsec-scale resolution and more successfully capture the multiphase ISM. Both major mergers and the accretion of cold gas via filaments are dominant contributors to a galaxy's total stellar budget and we examine SF at high resolution in both of these contexts. 

\end{abstract}

By performing adaptive mesh refinement (AMR) simulations in which we resolve the multiphase ISM and the formation of GMCs, we investigate the nature of galactic SF in two important contexts: major mergers (Section 1) and cold accretion via filaments at high-redshift (Section 2). 
               \firstsection
\section{Star formation modes during major mergers}

To date, we have performed $8$ simulations of $1$:$1$ galaxy mergers ($M_{\rm gal}\approx 8\times 10^{10}  {\rm M}_{\odot} $) with a maximum $5$pc spatial resolution and $1.6 \times 10^4$ ${\rm M}_{\odot}$ effective gas mass resolution using the AMR code {\sc ramses} \citep{ramses}, providing us with a representative sample that contains a variety of orbital parameters. To achieve our high resolution, we use a `pseudo-cooling' equation of state which dictates the temperature of gas based on its density \citep[see][for details]{antennae}. This approach has been shown to produce an ISM density power spectrum that agrees well with observations (Bournaud, 2010 in prep.). We choose a moderately high density threshold ($\sim10^3$ atoms ${\rm cm}^{-3}$) and a very low efficiency ($\sim 10^{-4}$) for the SF prescription, and employ the Sedov blastwave model of supernova (SN) feedback \citep{dubois_teyssier_sn} . This choice of parameters gives a star formation rate (SFR) of $\sim 1 {\rm M}_{\odot} {\rm yr}^{-1}$ in the galaxies when they are simulated in isolation, which is reasonable for a $z=0$ disc galaxy. Work is currently underway to test the sensitivity of the SFR to changes in these parameters.

\begin{figure}[h]
\centering
\begin{minipage}[b]{0.5\linewidth}
\centering
  \includegraphics[width=\textwidth]{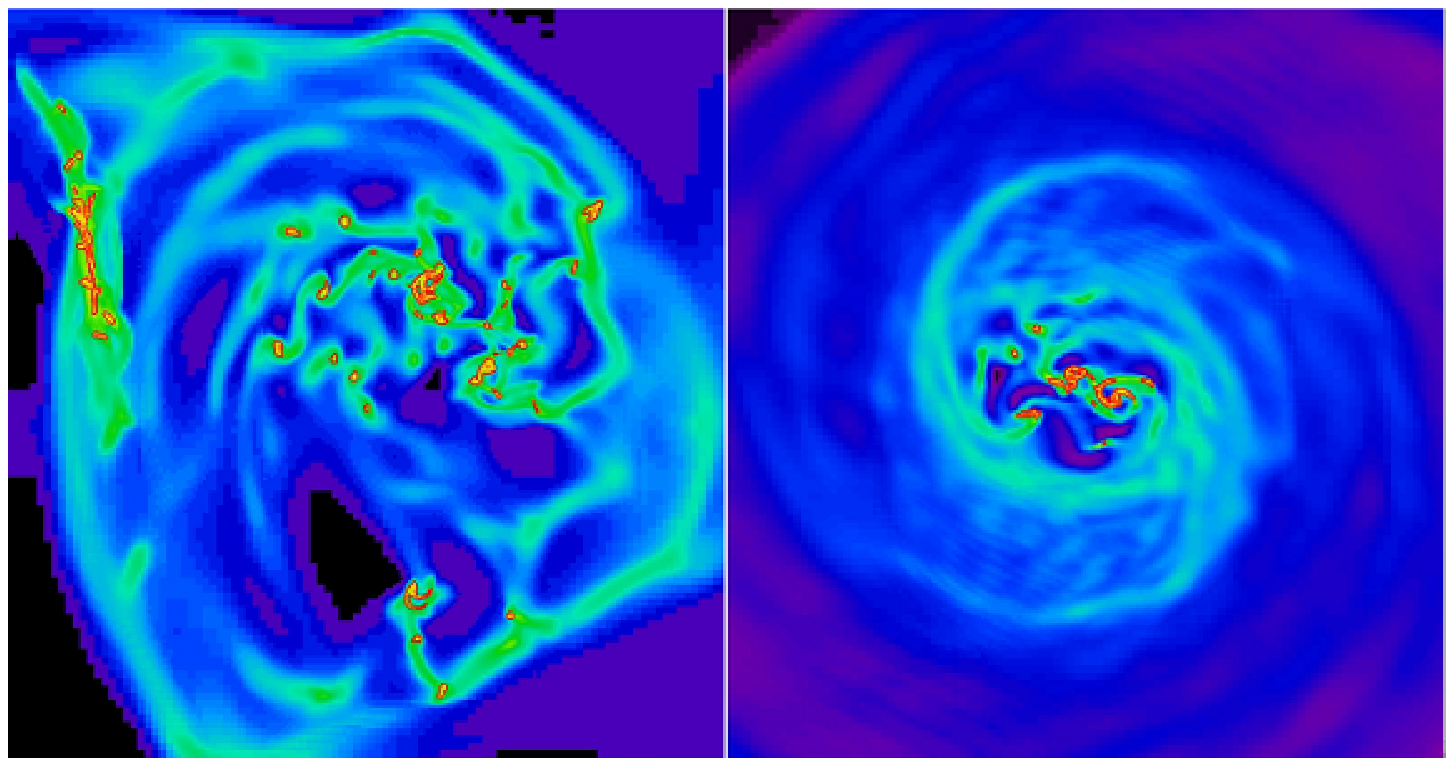}
\end{minipage}
\begin{minipage}[b]{0.3\linewidth}
\centering
   \includegraphics[width=\textwidth,trim = 20mm 8mm 15mm 15mm, clip]{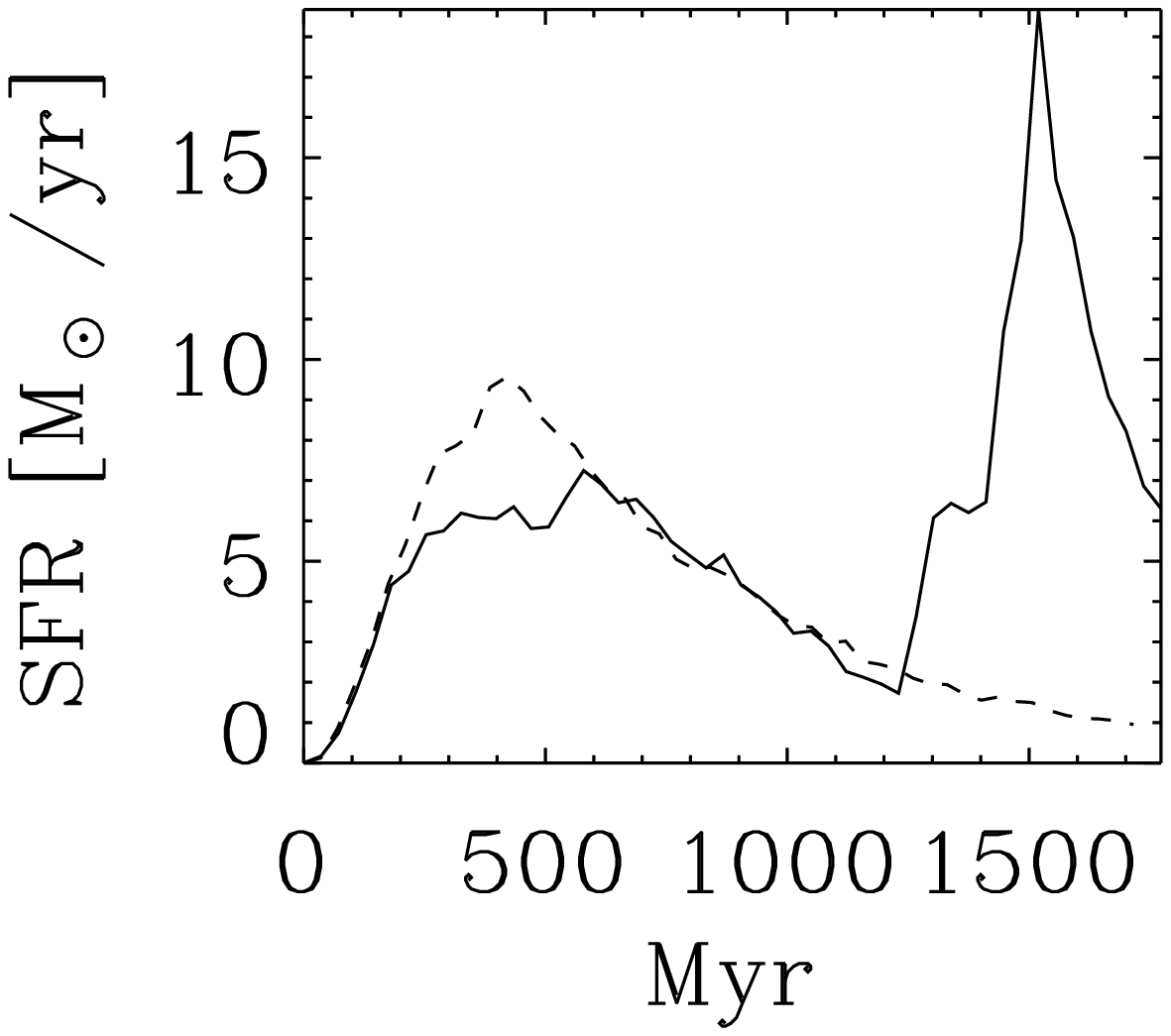}
\end{minipage}
 \caption{{\bf Left and Middle:} Gas density images of the merging galaxies, one face-on, one edge-on (left) and of one of the same galaxies when evolved in isolation (middle) at t=1.5 Gyr. Red contours indicate the regions with densities above the SF threshold. Size of image is $\approx 13{\rm kpc}$. {\bf Right:} SFR versus time for both of the merging galaxies (solid line) and the sum of the SFRs in the same two galaxies when evolved in isolation (dashed line).} 
\label{extended_sf}
\end{figure}

\subsection{Extended star formation}

In the classical picture, merger-induced starbursts result when tidal torques drive large amounts of gas into the centre of the merging structure, where it is compressed and forms stars rapidly (Struck, this meeting). We report the occurrence of another mode of merger-induced SF: clumpy, extended SF. In Fig.~\ref{extended_sf} we clearly demonstrate this alternative mode by comparing the gas density map of the merging galaxies (left) with that of one of the same galaxies when evolved in isolation (middle) at the same time instant ($t=1.5$ Gyr). Red contours indicate the high density, star-forming clumps and these are considerably more numerous and significantly less centrally concentrated in the merging galaxies (where the clumps are spread over a region of $\approx 7$kpc in diameter) than in the isolated galaxy (clumps spread over $\approx 3$kpc). This is particularly clear when comparing the merging galaxy seen face-on in the left-hand panel of Fig.~\ref{extended_sf} with the isolated galaxy (middle panel). By reference to the time evolution of the SFR (Fig.~\ref{extended_sf}, right), we can see that $t=1.5$ Gyr occurs just before the maximum SFR, indicating that this extended, clumpy SF is responsible for the starburst which has increased the SFR in the merger (solid line) by a factor of $\approx 10$ relative to the SFR in the same two galaxies evolved in isolation (dashed line) i.e. the starburst occurs {\it before} the final coalescence of the two galaxy centres.

This type of merger-induced star cluster formation over an extended region is also observed in the Antennae galaxies. \citet{antennae} have demonstrated that this is only reproduced in a simulation of the Antennae with {\sc ramses}  if a resolution of  $\sim 10$pc is achieved; they find that the star formation is no longer clustered if the gas is thermally supported (as occurs at lower resolutions), highlighting the strong dependence of the mode of  simulated SF on resolution.

\begin{figure}[h]
\centering
\begin{minipage}{0.75\linewidth}
  \includegraphics[width=0.48\textwidth]{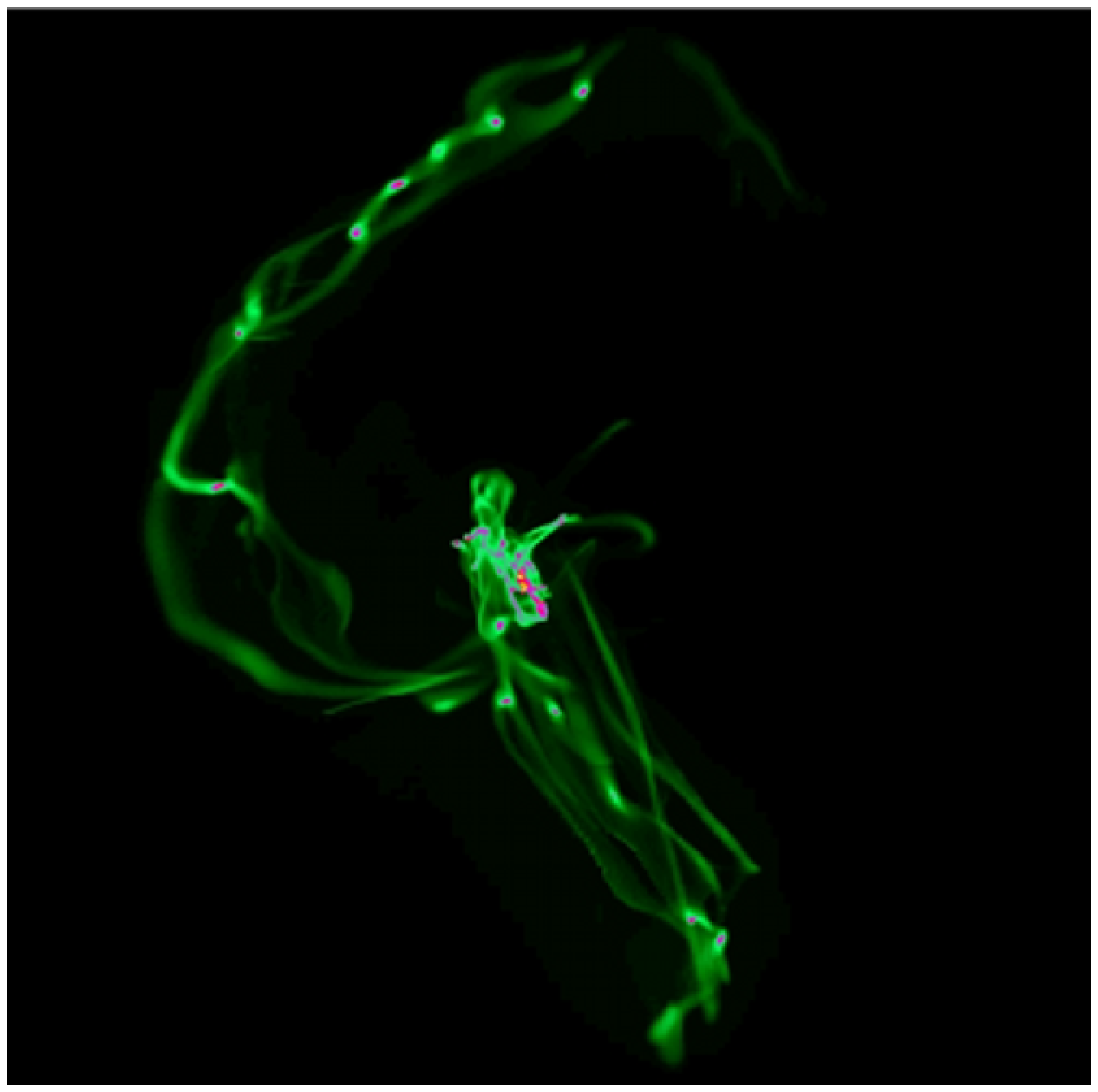}
   \includegraphics[width=0.5\textwidth,trim=20mm 0mm 20mm 10mm,clip]{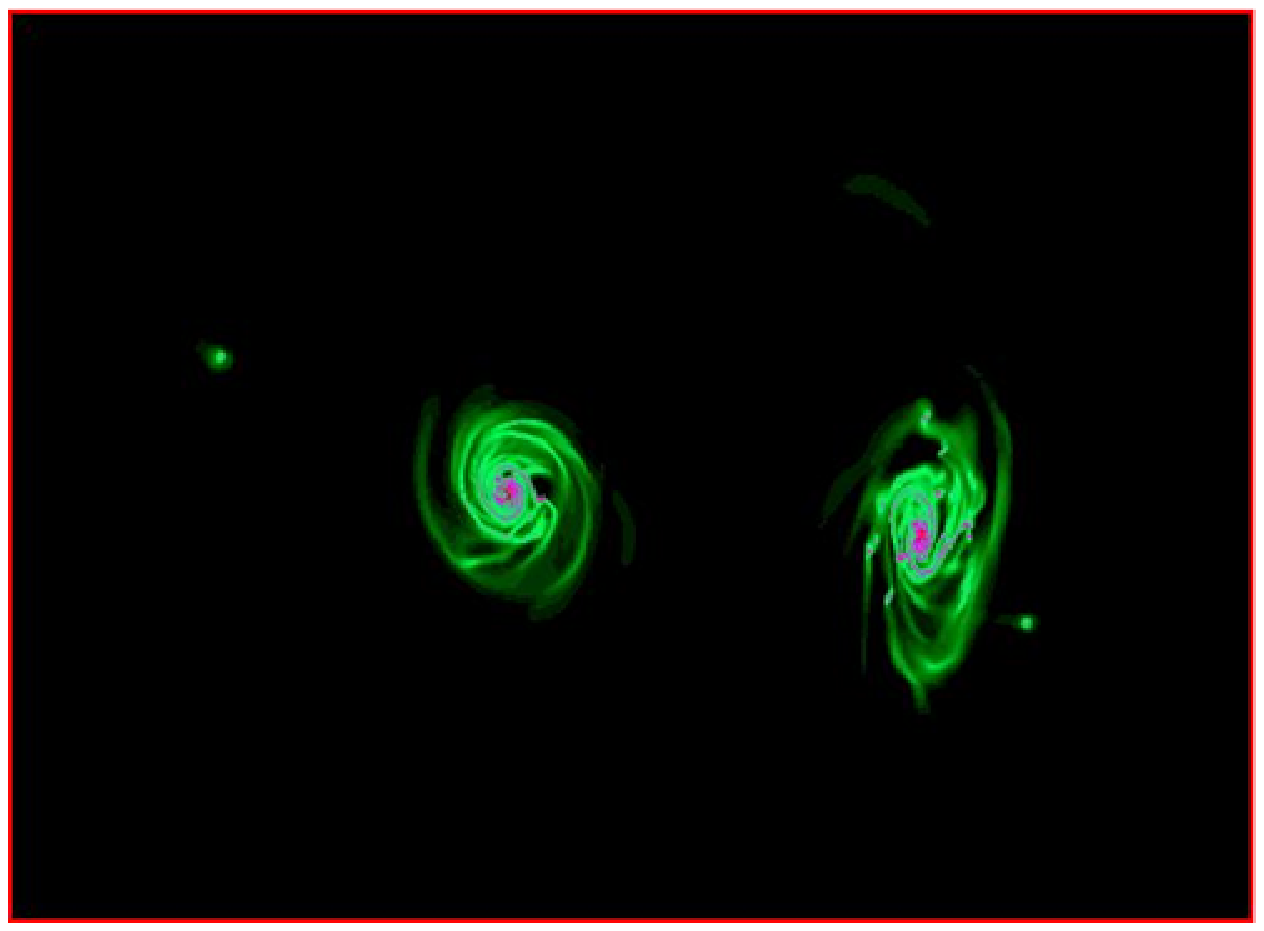}
   \end{minipage}
\begin{minipage}{0.23\linewidth}
 \caption{Gas density images from two different mergers. {\bf Left:} Formation of `beads on a string'  of length $\sim 100$ kpc. {\bf Right:} Two tidal dwarf galaxies (one with $M\sim 10^9 {\rm M}_{\odot}$) formed after the first pericenter passage.}
  \label{other_sf}
 \end{minipage}
\end{figure}

In Fig.~\ref{other_sf} we show some examples of the other varied star-forming features found serendipitously in our merger sample.

\section{Star formation during cold-mode accretion and the onset of a galactic wind}

The Nut Simulation  \footnote{Nut is the Egyptian goddess of the night sky} is an ultra-high resolution cosmological resimulation of a Milky Way like galaxy forming at the intersection of 3 filaments in a $\Lambda$CDM cosmology  \citep{adrianne_proc,julien_proc}. We perform the simulation with {\sc ramses} \citep{ramses} including cooling, a UV background \citep{uvbackground}, SF and SN feedback with metal enrichment \citep[][]{dubois_teyssier_sn}  achieving a maximum physical spatial resolution of $\approx 0.5$pc in the densest regions at all times.  A thin, rapidly rotating disc forms, which is continuously fed by cold gas that streams along the filaments at $\sim$Mach $5$. The disc is gravitationally unstable and fragments into star-forming clumps (see Devriendt et al (2010, in prep.) for details of the disc and Slyz et al (2010, in prep.) for a study of the clumpy star formation). SF in the galaxy, its satellites and haloes embedded in the filaments results in many SN explosions, which are individually resolved. The numerous bubbles overlap, creating an extended cavity through which the hot gas escapes the galactic potential in the form of a wind \citep[see][for figures illustrating this and additional technical details]{adrianne_proc,julien_proc}.  By $z=9$, when the dark matter halo mass has reached $\sim 10^9 {\rm M}_{\odot}$, the wind extends to $\approx 6  r_{\rm vir}$, filling the virial sphere with hot, diffuse gas not unlike the shock-heated gaseous halo found in galaxies above the mass threshold for a stable virial shock \citep[e.g.][]{shock_1d}. To date, the simulation has reached $z \approx7.5$ and it is still running. 

Here we present results from a study of the interaction between the cold filaments and the hot, far-reaching galactic wind and examine the impact this competition has on the SFR (Powell et al, 2010 in prep.).

\subsection{Inflows and outflows: measuring accretion and winds}

To assess the relative importance of the filaments and the galactic wind we measure the mass inflow and outflow rates within the virial radius of the galaxy.  We divide the gas into various phases with the use of temperature and density thresholds, similar to the approach in other simulation studies of hot and cold mode accretion \citep[e.g.][]{agertz_clumpygal}. Note, however, that the thresholds are resolution-dependent so ours differ from those in the previous lower resolution studies. The categories we define are: hot diffuse ($T > 2\times10^{5}$K, $n<10$ atoms ${\rm cm}^{-3}$),  warm diffuse ($2\times10^{4}$K $<T< 2\times10^{5}$K, $n<10$ atoms ${\rm cm}^{-3}$), cold diffuse ($T<2\times10^{4}$K, $n<0.1$ atoms ${\rm cm}^{-3}$), filaments  ($T<2\times10^{4}$K, $0.1\le n\le 10$ atoms ${\rm cm}^{-3}$) and `clumpy' ($n>10$ atoms ${\rm cm}^{-3}$).  We verify that these categories accurately separate the different gas components using 3D visualisation software.

\begin{figure}[ht]
\centering
  \includegraphics[width=0.29\textwidth,trim = 0mm 0mm 9mm 9mm, clip]{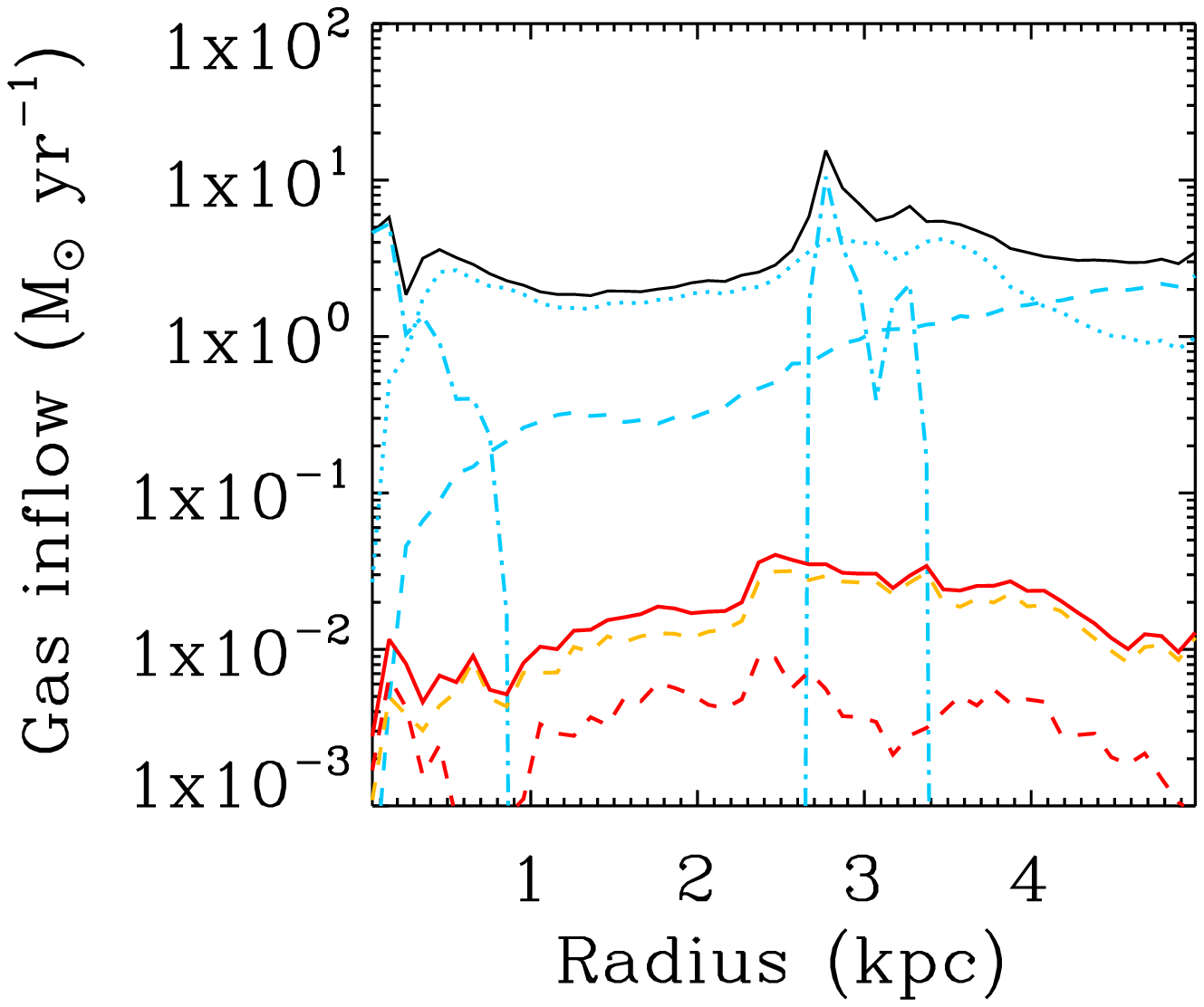}
   \includegraphics[width=0.29\textwidth,trim = 0mm 0mm 9mm 9mm, clip]{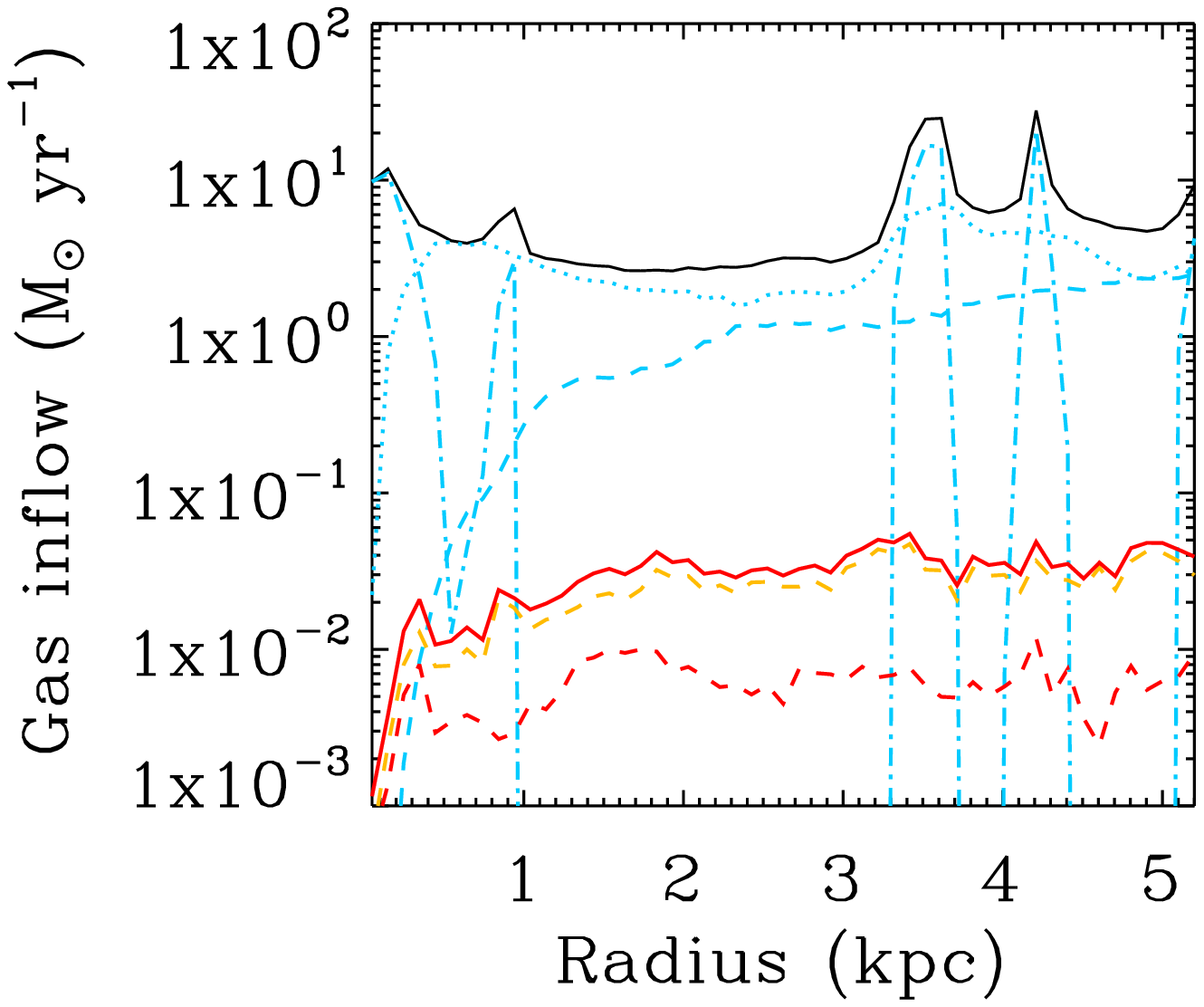}
   \includegraphics[width=0.29\textwidth,trim = 0mm 0mm 9mm 9mm, clip]{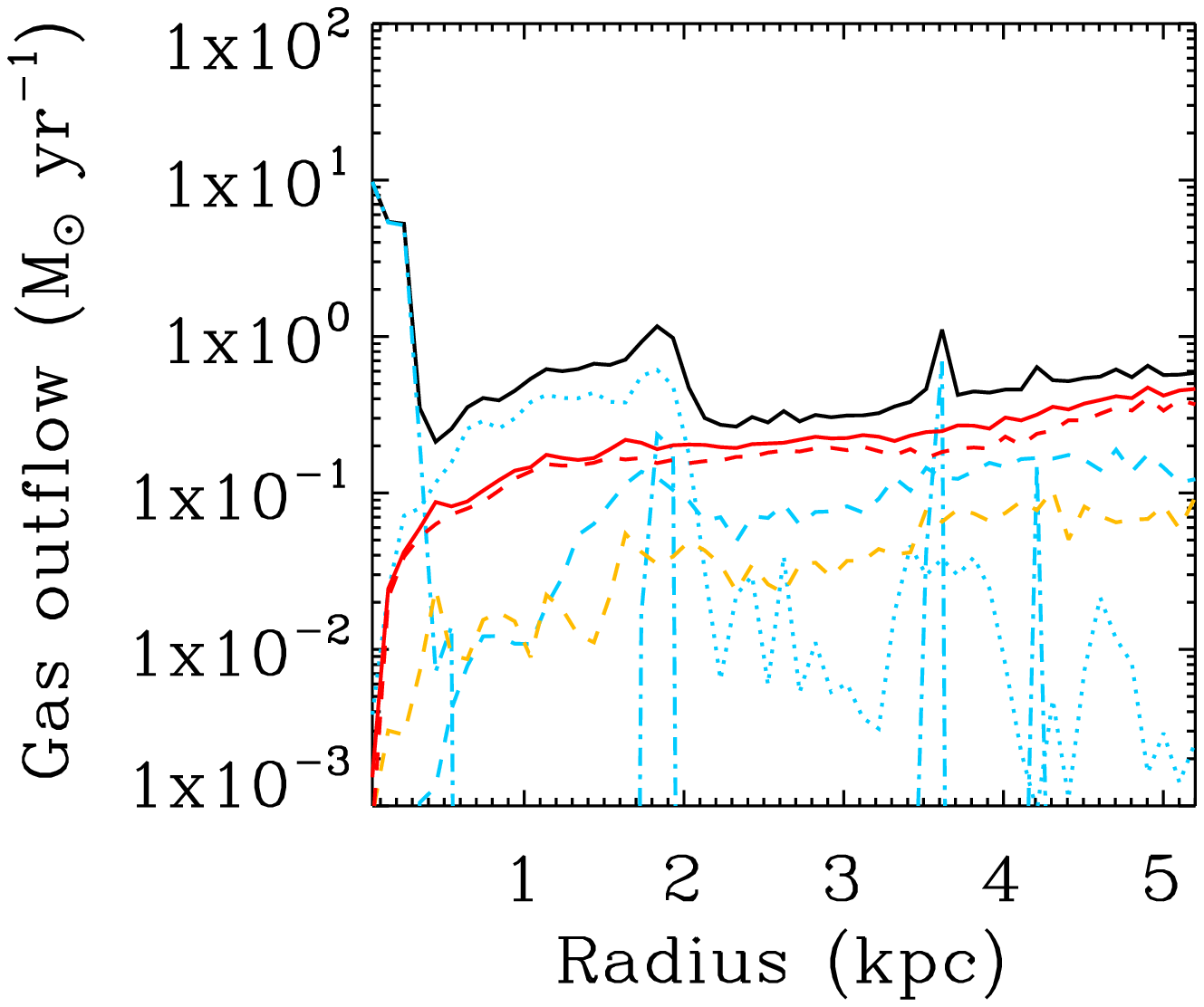}
 \caption{Gas inflow averaged in $100$ pc physical spherical shells out to $r_{\rm vir}$ in the control run without supernovae feedback (left) and the Nut simulation (middle) at $z=9$. Gas outflow in the Nut simulation (right). Flow rate of all gas (solid black line), subdivided into dense (blue dot-dash line), filamentary (blue dotted line), cold diffuse (blue dashed line), hot diffuse (red dashed line), warm diffuse (orange dashed line) and hot$+$warm diffuse (red solid line).}
\label{fluxvr}
\end{figure}

In Fig.~\ref{fluxvr} we compare the inflow in a control run with no SN feedback (left) with that in the Nut simulation (middle) and examine the outflow in the Nut simulation (right). We highlight several key results: 1) the inflow of cold gas in both runs (first two panels), is dominated at all radii by the contribution from the filaments (blue dotted line), which supply material right down to the disc,  2) the total inflow rate (first two panels, black solid line) is almost identical in the simulations with SN feedback (middle, the Nut simulation) and without SN feedback (left, control run), revealing that the galactic wind has had a negligible effect on the accretion rate, 3) the only significant mass outflow occurs in the Nut simulation (right) yet this is only around $10$ per cent of the mass inflow rate and 4) there is a non-negligible cold diffuse component (blue dashed line) in this outflow, which is gas that was previously inbetween the filaments that has been entrained by the wind.

\subsection{What happens to the star formation rate?}

We have established that the high-velocity, far-reaching galactic wind has been unable to impact the supply of cold gas, the `fuel' for SF, because it is delivered to the galaxy in supersonic, highly collimated, dense filaments. The other major influence on the SFR is the efficiency with which this fuel is converted into stars.

\begin{figure}[ht]
\centering
\begin{minipage}[b]{0.6\linewidth}
    \includegraphics[width=0.49\textwidth,trim = 8mm 8mm 10mm 5mm, clip]{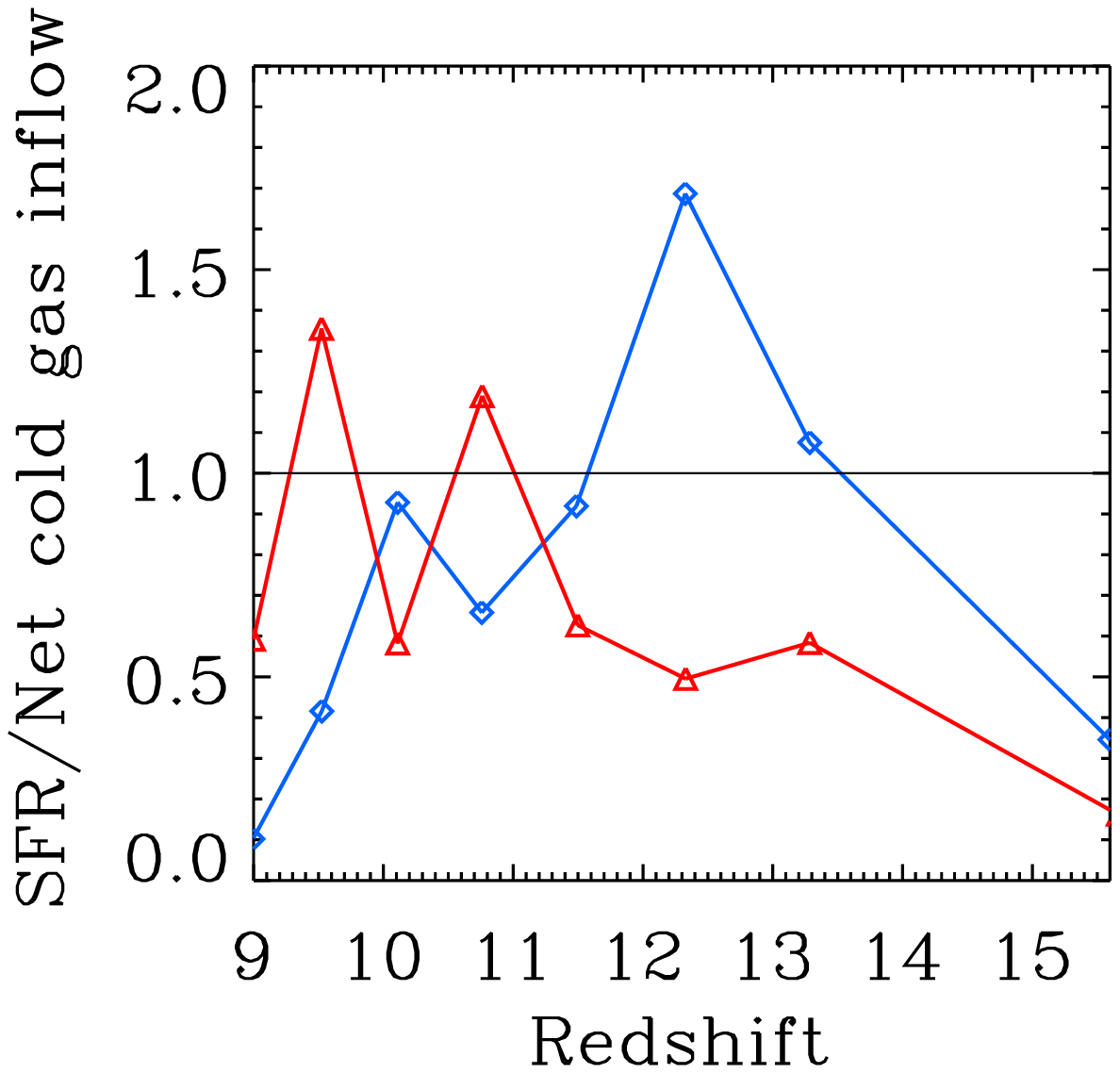}
 \includegraphics[width=0.49\textwidth,trim = 5mm 8mm 10mm 5mm, clip]{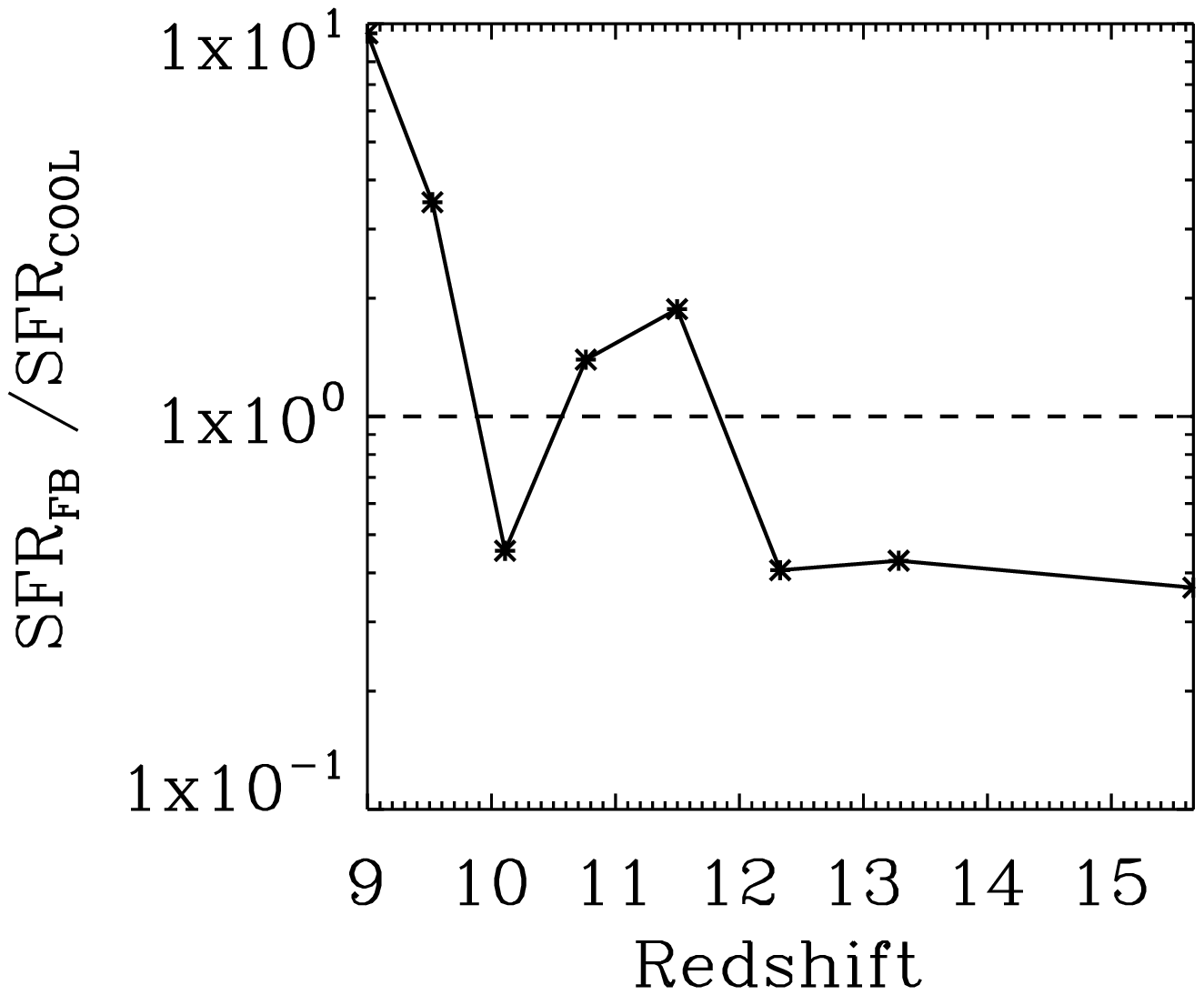}
 \end{minipage}
 \begin{minipage}[b]{0.31\linewidth}
    \caption{SFR of main galaxy averaged over 10Myr {\bf Left:} SFR divided by cold gas inflow rate for control run (blue) and Nut simulation (red).  {\bf Right:} SFR in Nut simulation divided by SFR in control run. Horizontal lines indicate when ratios $=1$.}
    \label{comparesfr}
     \end{minipage}
\end{figure}

The left-hand panel of Fig.~\ref{comparesfr} shows the ratio of the SFR over the net cold gas inflow rate for the control run (blue) and the Nut simulation (red). For both runs the ratio oscillates around $1$ (the horizontal line), indicating that the conversion of the filament-supplied cold gas to stars is very efficient. The mean efficiency in the central star-forming region is $\sim 0.1$ in both cases; an order of magnitude higher than the $1$ per cent efficiency used in {\sc ramses} to dictate the SFR on a cell by cell basis. The right-hand panel of  Fig.~\ref{comparesfr} shows the ratio of the SFR in the Nut Simulation to that in the control run (the actual SFR is a few ${\rm M}_{\odot} {\rm yr}^{-1}$). Approaching $z=9$, the SFR in the Nut simulation ({\it with} SN feedback) exceeds that in the control run ({\it without} SN feedback). We attribute this positive feedback effect to the metal enrichment in the Nut simulation (absent in the control run), which facilitates cooling of the gas, boosting SF.

\end{document}